%% file: main.tex
\documentclass[letterpaper, 10 pt, conference]{ieeeconf}

\IEEEoverridecommandlockouts 
                                                          
\input{packages}

\input{macros}

\title{\LARGE \bf Adaptive Shielding under Uncertainty}

\author{Stefan Pranger$^{1}$ \and  Bettina K\"onighofer$^{1,2}$ \and  Martin Tappler$^{4}$ \and 
Martin Deixelberger$^{1}$ \and Nils Jansen$^{3}$ \and Roderick Bloem$^{1}$ 
\thanks{*This work was supported by}
\thanks{$^{1}$
Graz University of Technology, Institute IAIK, Graz, Austria
}
\thanks{$^{2}$
Silicon Austria Labs, TU-Graz SAL DES Lab, Graz, Austria
}
\thanks{$^{3}$
Radboud University Nijmegen, Nijmegen, The Netherlands
}
\thanks{$^{4}$
Schaffhausen Institute of Technology, Schaffhausen, Switzerland
}
}

\begin{document}
\maketitle

\begin{abstract}
	\input{abstract}
\end{abstract}

\section{Introduction}\label{sec:intro}
\input{introduction}

\section{Preliminaries}\label{sec:prelim}
\input{preliminaries}

\section{Adaptive Shielding Setting} \label{sec:setting}
\input{setting}

\section{Adaptive Shielding Framework} \label{sec:method}
\input{method}

\section{Implementation and Experiments}\label{sec:experiments}

\input{experiments}


\section{Conclusion and Future Work}\label{sec:conclusion}

\input{conclusion}



\bibliographystyle{IEEEtran}
\bibliography{IEEEabrv,literature}
\end{document}

%% file: packages.tex
\usepackage{amsmath,amssymb,amsfonts}
\usepackage{mathtools}

\usepackage{amsthm}

\usepackage[usenames,dvipsnames]{color}
\usepackage{xcolor}
\usepackage{xspace}
\usepackage{microtype}
\usepackage{todonotes}
\usepackage{colonequals}
\usepackage{wasysym}
 \usepackage{booktabs}
\usepackage[capitalize]{cleveref}

\usepackage{multirow}
\usepackage{multicol}

\usepackage{paralist}

\usepackage{subfigure}
\usepackage{url}
\usepackage{graphicx}
\usepackage{float}
\usepackage{algorithm}
\usepackage{listings}
\usepackage{nicefrac}
\usepackage{adjustbox}
\usepackage{tikz}
\usetikzlibrary{calc,positioning}
\usepackage{pgfplots}
\usepackage{pgfplotstable}
\usepgfplotslibrary{groupplots}
\pgfplotsset{compat=newest}

\usepackage{wrapfig}

\usetikzlibrary{arrows,decorations.pathmorphing,positioning,fit,trees,shapes,shadows,automata,calc}

%% file: macros.tex
\newcommand\shieldsym[1][]{%
\raisebox{-1.5pt}{\tikzset{
    shield width/.store in=\shieldwidth,
    shield width=1.5ex,
    shield height/.store in=\shieldheight,
    shield height=1.75ex
}%
\tikz [baseline,#1] \draw (0,\shieldheight) -- (0,\shieldwidth/2) arc [radius=\shieldwidth/2, start angle=-180, end angle=0] -- (\shieldwidth,\shieldheight) -- cycle;%
}}

\newtheorem{definition}{Definition}

\newcommand{\tinyshield}{\text{\tiny\shieldsym}}
\newcommand{\shielded}[1]{\ensuremath{#1_{\tinyshield}}}

\newcommand{\Distr}{\mathit{Distr}}

\newcommand{\R}{\mathbb{R}}

\newcommand{\N}{\mathbb{N}}

\newcommand{\sinit}{s_0} 
\newcommand{\states}[1][]{\mathcal{S}_{#1}}

\newcommand{\dtmc}{\mathcal{D}}
\newcommand{\DtmcInit}[1][]{\ensuremath{\dtmc{#1}=(\states{#1},\sinit{#1},\pmdp{#1})}}

\newcommand{\mdp}{\mathcal{M}}

\newcommand{\MdpTupleR}[1][]{\ensuremath{(S{#1},\sinit{#1},\Act{#1},\pmdp{#1})}}
\newcommand{\MdpInitR}[1][]{\ensuremath{\mdp{#1}=\MdpTupleR[#1]}}
\newcommand{\MdpTupleShielded}[1][]{\ensuremath{(\shielded{\states{#1}},\shielded{\sinit{#1}},\shielded{\Act{#1}},\shielded{\pmdp{#1}})}}

\newcommand{\act}[1][a]{\alpha} 
\newcommand{\Act}{\mathcal{A}}

\newcommand{\e}{\ensuremath{\mathbb{E}}}

\newcommand{\costFunction}{c}

\newcommand{\pmdp}{\mathcal{P}}

\newcommand{\ltp}[1][L]{\mathcal{L}}   

\renewcommand{\path}{\pi}

 \newcommand{\twodots}{\mathinner {\ldotp \ldotp}}

\newcommand{\bddstate}[1]{s_{\mathbb{B}}{}}

\newcommand{\tool}[1]{\texttt{#1}\xspace}

\newcommand{\storm}{\tool{Storm}}

\newcommand{\sumo}{\tool{SUMO}}

\DeclareRobustCommand{\cpp}
{\valign{\vfil\hbox{##}\vfil\cr
   \textsf{C\kern-.1em}\cr
   $\hbox{\fontsize{\sf@size}{0}\textbf{+\kern-0.05em+}}$\cr}%
}

\selectcolormodel{cmy}

\definecolor{lgrey}{rgb}{0.8,0.8,0.8}
\definecolor{grey}{rgb}{0.5,0.5,0.5}

\definecolor{lightblue}{rgb}{0.8,0.8,1.0}
\definecolor{lightred}{rgb}{1.0,0.8,0.8}
\definecolor{lightgreen}{rgb}{0.8,1.0,0.8}
\definecolor{angrygreen}{cmyk}{0.279,0,0.91,0.08}
\definecolor{lightred}{rgb}{1.0,0.8,0.8}
\definecolor{pink}{rgb}{1.0,0.1,1.0}

\definecolor{prismgreen}{HTML}{009900}
\definecolor{prismred}{HTML}{cc0000}
\definecolor{prismblue}{HTML}{0000cc}

\lstdefinelanguage{Prism}{
        basicstyle=\color{prismred}\scriptsize\ttfamily,
        literate=*	{0}{{\textcolor{prismblue}{0}}}{1}
			{1}{{\textcolor{prismblue}{1}}}{1}
			{2}{{\textcolor{prismblue}{2}}}{1}
			{3}{{\textcolor{prismblue}{3}}}{1}
			{4}{{\textcolor{prismblue}{4}}}{1}
			{5}{{\textcolor{prismblue}{5}}}{1}
			{6}{{\textcolor{prismblue}{6}}}{1}
			{7}{{\textcolor{prismblue}{7}}}{1}
			{8}{{\textcolor{prismblue}{8}}}{1}
			{9}{{\textcolor{prismblue}{9}}}{1}
			{.0}{{\textcolor{prismblue}{.0}}}{2}
			{.1}{{\textcolor{prismblue}{.1}}}{2}
			{.2}{{\textcolor{prismblue}{.2}}}{2}
			{.3}{{\textcolor{prismblue}{.3}}}{2}
			{.4}{{\textcolor{prismblue}{.4}}}{2}
			{.5}{{\textcolor{prismblue}{.5}}}{2}
			{.6}{{\textcolor{prismblue}{.6}}}{2}
			{.7}{{\textcolor{prismblue}{.7}}}{2}
			{.8}{{\textcolor{prismblue}{.8}}}{2}
			{.9}{{\textcolor{prismblue}{.9}}}{2}
			{[}{{\textcolor{black}{[}}}{1}
			{]}{{\textcolor{black}{]}}}{1}
			{+}{{\textcolor{black}{+}}}{1}
			{-}{{\textcolor{black}{-}}}{1}
			{=}{{\textcolor{black}{=}}}{1}
			{<}{{\textcolor{black}{<}}}{1}
			{>}{{\textcolor{black}{>}}}{1}
			{\&}{{\textcolor{black}{\&}}}{1}
			{|}{{\textcolor{black}{|}}}{1}
			{:}{{\textcolor{black}{:}}}{1}
			{;}{{\textcolor{black}{;}}}{1}
			{(}{{\textcolor{black}{(}}}{1}
			{)}{{\textcolor{black}{)}}}{1}
			{..}{{\textcolor{black}{..}}}{2},
        keywords= {bool,C,ceil,const,ctmc,double,dtmc,endinit,endmodule,endrewards, endsystem,F,false,floor,formula,G,global,I,init,int,label,max,mdp,min,module,nondeterministic,P,Pmin,Pmax,prob,probabilistic,R,rate,rewards,Rmin,Rmax,S,stochastic,system,true,U,X},
        keywordstyle={\bfseries\color{black}},
        numberstyle=\footnotesize\color{black},
        comment=[l] {//}, morecomment=[s]{/*}{*/},
        commentstyle= \color{prismgreen},
        tabsize=4,
        captionpos=b,
        escapechar=@
}

%% file: abstract.tex
This paper targets control problems that exhibit specific safety and performance requirements.
In particular, the aim is to ensure that an agent, operating under uncertainty, will at runtime strictly adhere to such requirements.
Previous works create so-called shields that correct an existing controller for the agent if it is about to take unbearable safety risks.
However, so far, shields do not consider that an environment may not be fully known in advance and may evolve for complex control and learning tasks.
We propose a new method for the efficient computation of a shield that is adaptive to a changing environment. 
In particular, we base our method on problems that are sufficiently captured by potentially infinite Markov decision processes (MDP) and quantitative specifications such as mean payoff objectives.
The shield is independent of the controller, which may, for instance, take the form of a high-performing reinforcement learning agent.
At runtime, our method builds an internal abstract representation of the MDP and constantly adapts this abstraction and the shield based on observations from the environment.
We showcase the applicability of our method via an urban traffic control problem.

%% file: introduction.tex
Nowadays controllers 
are increasingly sophisticated and complex and make extensive use of machine learning, such as
reinforcement learning, and other optimization techniques for smart control in
open environments~\cite{FultonP18}. However, due to the complexity of such controllers, it is practically infeasible to cover the entire input space of a system with test cases to guarantee safety or any level of performance.
Learned controllers introduce additional challenges.
In particular, it is difficult to achieve acceptable performance
when encountering untrained scenarios and to add new features to the controller without retraining or inducing negative side effects.

\textbf{Shielding.} One suitable technique that delivers theoretical guarantees w.r.t. qualitative or quantitative objectives at runtime are so-called shields~\cite{AvniBCHKP19,BloemKKW15,AlshiekhBEKNT18}.
In this paper, we consider shields that enforce \emph{quantitative} measures~\cite{AvniBCHKP19}.
At each time point, the shield reads the command issued by the controller and
may choose to alter it before passing it on to the environment.

Shields should always be lightweight, we assume that the controller
generally performs well.
When combining the shield and the controller, intuitively, the
controller should be active most of the time, and the shield intervenes only when required. 
Note that if the controller and the shield have different yet strong objectives, any interference of the shield may cause a performance drop in the controller's objective.
On the other hand, if the shield and the controller have similar performance measures, the shield may even improve the controller's performance by interfering (especially in unusual situations).

We base the decision whether the shield should interfere at any point in time on \textit{quantitative measures}. 
In particular, we formalize a \emph{performance objective} to be minimized by the shield, and an \emph{interference cost} for changing the output of the controller. 
Intuitively, the shield needs to balance the cost of interfering with the decrease in performance of not interfering. 

Our synthesis procedure assumes a stochastic environment, which includes the controller. 
It captures the task of shielding in an MDP, where actions denote changing the controller output. A policy for choosing actions is a concrete shield in this interpretation. We compute shields that maximize a single  mean  payoff  objective  obtained  from  combining the performance and the interference measure, thus guaranteeing maximal performance
with minimal interference.

\textbf{Need for adaptivity.}
The computation of shields necessitates a faithful abstraction of the physical environment dynamics.
In case of an inaccurate model, the interference of the shield may be unjustified, which can result in drops for both objectives.
Even if the model was accurate at the beginning, in real-world scenarios the environment in which the shield operates may change over time.
Therefore, the shield has to adapt to the new situation and needs to
adjust its environment model. 

\textbf{Idea of adaptive shielding.}
In this work, we aim to address the problem of inaccurate modeling by
proposing an approach based on model refinement and online estimation of transition probabilities. 
Initially, a first shield is synthesized from an initial abstract model of the environment (including the controller). 
During run-time, the shield is applied.
Additionally, a monitor observes the behavior of the environment and the controller. 
Every $t$ time units, the synthesis procedure updates the abstract model and computes a new shield. The new shield replaces the previous one during further operation.

\textbf{Updating the models.}
In this paper, we implement effective and efficient shielding based on two models capturing our observations of the environment. 

\noindent
+{1) \em Infinite-state MDP $\mdp$:} During run-time, we continuously update an infinite-state MDP $\mdp$ using our knowledge of the whole environment. 
    We base an online estimation of the transition probabilities of $\mdp$ on observations of the environment.

\noindent    
 {2) \em Abstracted finite-state MDP $\shielded{\mdp}$:} An abstraction $\shielded{\mdp}$ of $\mdp$ allows us to efficiently compute shields. Whenever we deduce that our current abstraction $\shielded{\mdp}$ is too coarse, we refine it by taking additional parts of $\mdp$'s state space into account.
    We also regularly update $\shielded{\mdp}$ to ensure that changes of $\mdp$'s transition function are reflected in $\shielded{\mdp}$. 

Continuous updates of the probabilistic transition function of $\mdp$ and $\shielded{\mdp}$ ensure that derived shields perform well in a constantly changing environment.

\textbf{Generic shields due to adaptivity.}
An additional advantage of our adaptive shielding framework is the possibility to synthesize generic shields and apply them in different concrete settings. 
Here, initially we assume a generic environment where all environment changes, that is, transitions of $\mdp$, are equally likely. From this model, we synthesize a generic shield which can be applied in several concrete settings, and the shields adapt over time.

\textbf{Case study: Urban Traffic Control.} We demonstrate the applicability of our framework by shielding traffic-light controllers in various road networks. In such a setting, an accident on the highway causes drastic changes in the traffic density on smaller roads, resulting in a traffic jam when using an unshielded traffic light controller. Due to the adaptive nature of the shielding approach, the shielded controller is able to maintain the traffic flow through the city.


\textbf{Related Work.}
The general set of techniques that ensure the correctness of a controller at run-time is referred to as run-time enforcement (RE)~\cite{FalconeP19,RenardFRJM19}.
The concept of a correct-by-construction \emph{safety-shield} to enforce such correctness with respect to a temporal logic specification, with the additional goal of minimal interference with the controller's output, was proposed first in~\cite{BloemKKW15,AlshiekhBEKNT18}.
Several extensions exist~\cite{DBLP:conf/amcc/BharadwajBDKT19,wu2019shield}.
Closest to our approach is \cite{AvniBCHKP19} which proposed \emph{optimal-shields} that enforce 
quantitative objectives at run-time.
We extend this work by dealing with the consequences of an incorrect or incomplete model that is used for the computation of the shield. 
In particular, our adaptive shielding technique uses abstraction refinement and online probability estimation to deal with incorrect models and to adjust to new situations. 
Furthermore, \cite{AvniBCHKP19} treated the controller adversarially. 
In our  approach, we make initial assumptions on the behavior of the controller and update them during run-time. 

Through online estimation of transition probabilities, we basically learn an MDP model on-the-fly during run-time. Active automata learning techniques for MDPs~\cite{DBLP:conf/icmla/ChenN12,DBLP:journals/ml/MaoCJNLN16,DBLP:conf/fm/TapplerA0EL19} also estimate transition probabilities on-the-fly and learn a structure in parallel.
While we assume a known structure, we consider a larger state-space and we learn from a single prefix of a path, whereas automata learning techniques usually sample many finite paths.

Finally, such a shield for MDPs is relevant for the area of safe reinforcement learning~\cite{garcia2015comprehensive,DBLP:conf/aaai/FultonP18}.
Safe reinforcement learning via shielding has been considered in~\cite{AlshiekhBEKNT18,DBLP:conf/concur/0001KJSB20,DBLP:journals/corr/abs-1904-07189}, but none of these approaches takes changing environments into account.
Finally, there are approaches that guide reinforcement learning via temporal logic constraints towards the verification of MDPs~\cite{DBLP:conf/atal/HasanbeigAK20,DBLP:conf/tacas/HahnPSSTW19}.

%% file: preliminaries.tex
In this section, we introduce the required models and properties considered in this paper.

We use $1$-based indexed access on tuples which we denote by $t[i]$, where $t$ is a tuple and $i$ is an index. The notation $t[I \leftarrow k]$ denotes overwriting of the values in a tuple $t$  at indexes given by the set $I$ with values given by a function $k$, that is, $(t[I \leftarrow k])[i] = k(i)$ for $i \in I$ and $(t[I \leftarrow k])[i] = t[i]$ for $i\notin I$.

A \textbf{word} is a finite or infinite sequence of elements from some alphabet $\Sigma$. The set of finite words over $\Sigma$ is denoted $\Sigma^*$, and the set of infinite words over $\Sigma$ is written as $\Sigma^\omega$. The union of $\Sigma^*$ and $\Sigma^\omega$ is denoted by the symbol $\Sigma^\infty$.

A \textbf{probability distribution} over a {(finite)} set $X$ is a function $\mu: X \rightarrow [0, 1] \subseteq \R$ with $\sum_{x\in X}\mu(x) = \mu(X) = 1$.
The set of all distributions on $X$ is denoted by $Distr(X)$. 

A \textbf{Markov decision process} (MDP) $\MdpInitR$ is a tuple
with a finite set $\states$ of states, a unique initial state $s_0 \in \states$, a finite set $\Act=\{a_1\dots, a_n\}$ of actions, and
a \emph{probabilistic transition function} $\pmdp: \states \times \Act \rightarrow Distr (\states)$.
For all $s \in \states$ the available actions are $\Act(s) = \{a \in \Act | \pmdp(s, a) \neq \bot\}$ and we assume $|\Act(s)| \geq 1$.
A \emph{path} in an MDP $\mdp$ is a finite (or infinite) sequence $\rho=s_0a_0s_1a_1\ldots$ with $\pmdp(s_i, a_i, s_{i+1}) > 0$ for all $i\geq 0$.

Non-deterministic choices in an MDP are resolved by a
so-called \emph{policy}. 
For the properties that we consider in this paper, memoryless deterministic policies are sufficient~\cite{BK08}. These are functions $\pi : \states \rightarrow \Act$ with $\pi(s) \in \Act(s)$.
We denote the set of all memoryless deterministic policies of an MDP by $\Pi$.
Applying a policy $\pi$ to an MDP yields an induced \textbf{Markov chain} (MC) \DtmcInit ~with $\pmdp : \states\rightarrow\Distr(\states)$ where all nondeterminism is resolved.

A \emph{cost function}  $\costFunction: \states \times \Act \rightarrow\R_{\geq 0}$ for an MDP $\mdp$ adds a cost  to every state $s$ and action $a$ enabled in $s$.
For an infinite path $\rho = s_0 a_0 s_1 a_1 \dots$, the cost function
$c(s_i,a_i)$ returns the cost for every transition at step $i\geq0$. 



\textbf{Mean-payoff objectives~\cite{ChatterjeeKK17,BrazdilCFK15}.}
Given a strategy $\pi$, the \emph{$n$-step average cost} then is 
\begin{align*}
v_n^\pi(s)&=\e^\pi_s\bigg[\frac{1}{n} \sum_{i=0}^{n-1} \costFunction(s_i,a_i)\bigg],\\
\intertext{and the \emph{long-run average cost} of the strategy $\pi$ is }
v^\pi(s) &= \limsup\limits_{n\rightarrow \infty} v_n^\pi(s).
\end{align*}
For finite MDPs, the optimal limit-superior (also called the \emph{value}) is obtained by some memoryless deterministic strategy $\pi^\star\in\Pi$.
$$v^\star(s)=\min_{\pi\in\Pi} v^\pi(s)=\lim_{n\rightarrow \infty} v^{\pi^\star}.$$

In this paper, we compute an approximate solution for the mean payoff optimization problem with two objectives by computing the solution 
for a single mean-payoff objective, where every time the objective is obtained as a weighted sum of the objectives for which the Pareto curve is generated. The weights are selected in a way similar to~\cite{AvniBCHKP19}, allowing us to obtain the approximation of the curve.
Given two cost functions $c_1$ and $c_2$ and a factor $\gamma \in [0, 1]$ with which we weigh the two costs, we compute the \emph{$n$-step average weighted cost} with
$$v_n^\pi(s)=\e^\pi_s\bigg[\frac{1}{n} \sum_{i=0}^{n-1} (\gamma \cdot \costFunction_1(s_i,a_i) + (1-\gamma) \cdot \costFunction_2(s_i,a_i) )\bigg].$$

%% file: setting.tex

\subsection{Quantitative Shielding} 
We consider a stochastic environment that includes a global controller (obtained via RL) that issues local commands, or various local controllers which collaborate to optimally achieve some global goal. 
The shielding framework proposes to attach \emph{local shields} to alter local commands. 
\emph{Hence, we synthesize each local shield individually w.r.t. the environment and a local controller.}
A shield serves as a proxy between the controller and
the environment. At each point in time, the controller reads the state of the environment
and issues a command. Rather than directly feeding the command to the environment, the
shield first reads it along with an abstract state of the environment. The shield can then choose to
keep the controller’s command or alter it, before issuing the command to the environment. 
A shield minimizes two quantitative run-time measures: the \emph{shield performance} objective
and the \emph{interference} costs. When combining the shield and the controller, intuitively, the controller should be active for the majority of the time and the shield intervenes only when required; i.e.,
the shield needs to balance the cost of interfering with the decrease in the shield's performance of not
interfering. 

\textbf{Quantitative Shield Synthesis.}
We are given an abstraction MDP $\shielded{\mdp}$ with two cost functions: 
$c_1$ denotes the performance objective of the shield, and $c_2$ denotes the costs for interference. A factor $\gamma \in [0, 1]$ weighs the two scores.
A quantitative shield is computed by solving the MDP $\shielded{\mdp}$ w.r.t. the with $\gamma$ weighted mean payoff objective of $c_1$ and $c_2$ and implements an optimal policy $\pi^\star\in \Pi$ 
that achieves an optimal value $v^\star$.

\emph{\emph{\textbf{Example.}} As a running example, we consider a global traffic light controller 
that \emph{minimizes the total waiting time of all cars in the city}.
A local shield overwrites the command of the controller of a single junction.
The shield performance objective could supplement the objective of the controller, e.g., by balancing the number of waiting cars per incomming road. In unforeseen scenarios, e.g., with unusually dense traffic flow in one direction, the shield may improve the total waiting time (the controller's objective) by interfering.
A simple example of an interference score charges the shield $1$ for every change of action and charges $0$ when no change is made.
}

\subsection{Models} 
\textbf{Infinite-state model $\mdp$.}
We consider an environment including a controller, for which we assume stochastic behavior.
The stochastic behavior determines all actions of the controller via probabilities, thus the only non-determinism occurs in the actions of the shield. 
Combining these models yields an MDP $\MdpInitR$.
We assume that each state $s\in \states$ of $\mdp$ is a tuple of $n$ discrete variables $s=(v_1,\dots,v_n)$. Each variable $v_i$ is associated with a domain  of possible values $D_{vi}\subseteq \N$, i.e., $\states = D_{v1} \times \dots \times D_{vn}$. 
The state space $\states$ for modeling the environment may be infinite, but it needs to be countable. Furthermore, we assume that for each state-action $(s,a)$ the support of $\pmdp(s,a)$ is finite, that is, there are finitely many possible successor states following $(s,a)$. 

\emph{\emph{\textbf{Example.}} 
In our example of urban traffic control, the dimensions of the state space of $\mdp$ represent the number of cars waiting per road and the current command of the controller, i.e.,
states have the form $(n,s,e,w,ctr)$, see Fig.~$1$. 
The value $ctr$ denotes the current command of the controller, for instance, $\mathit{NS_c}$ denotes that the north-south direction should get the green light next. 
Hence, $ctr$ can take finitely many values, whereas the other values are theoretically unbounded.
The shield decides the next setting of the traffic light. It can either give the same command as the controller (here $\shielded{\mathit{NS}}$) or decide to deviate (here $\shielded{\mathit{EW}}$).
Upon issuing the action, the probabilities capture
the likeliness of the next queue sizes (influenced by the assumed distribution of incoming cars and the current traffic light setting commanded by the shield) combined with the chance of the next command of the controller.
The assumption of finite successor states is fulfilled, as the number of cars approaching a traffic light in a single time step is limited.  
}
\begin{figure}[ht]
\center
\includegraphics[height=3cm]{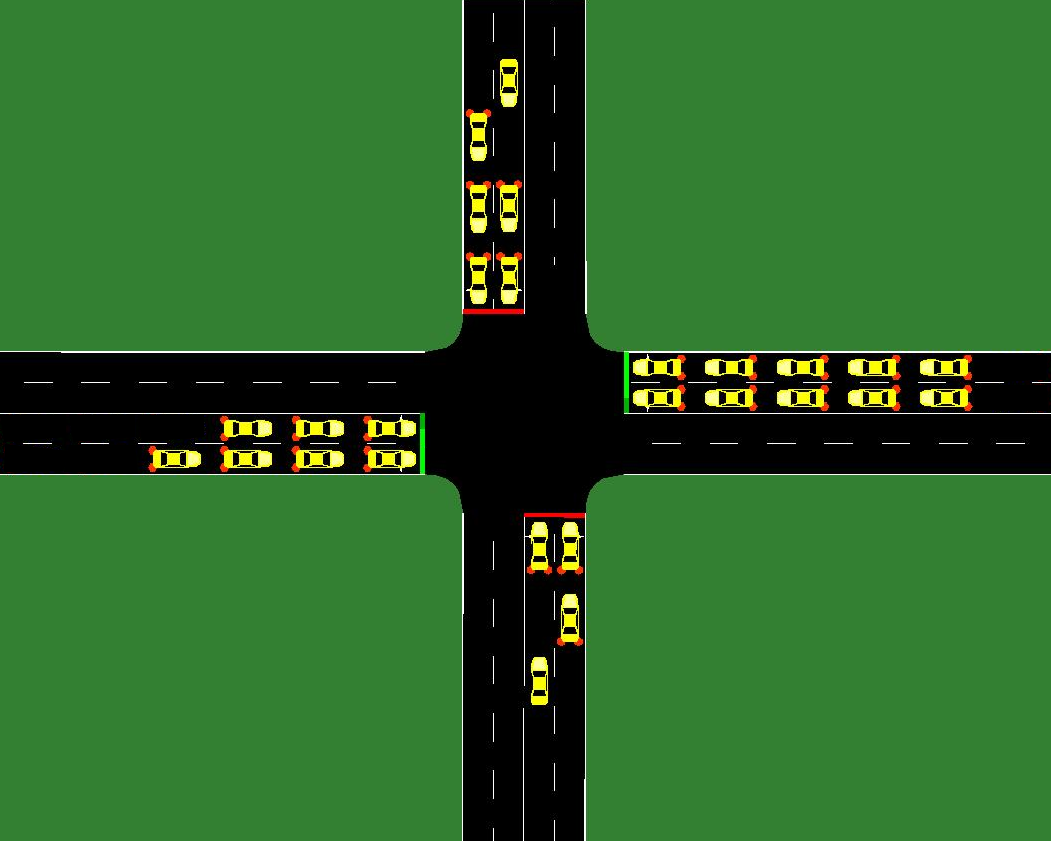}
\includegraphics[height=2cm]{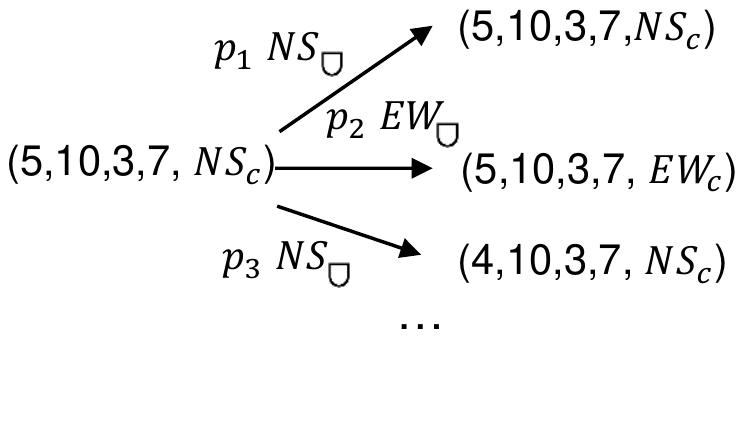}
\vspace{-0.2cm}
\caption{On the left, a concrete state depicted in the traffic simulator SUMO. On the right, we depict the corresponding state of $\mdp$ and some outgoing transitions. 
}
\label{fig:states_M}
\end{figure}

\textbf{Finite-state abstraction $\shielded{\mdp}$ of $\mdp$.}
We use  a finite-state abstraction $\shielded{\mdp}$ as underlying model to synthesize the shield. 
The idea is that $\shielded{\mdp}$ and $\mdp$ are equivalent on regularly visited states of the state space,
and rarely visited states are basically abstracted away. 

We define the abstraction MDP $\shielded{\mdp}$
by limiting the values that variables $v_i\in s$ can take. A cut-off function $k$ defines the maximal value for each variable $v_i \in s$, i.e., $v_1 \leq k(1), \dots , v_n \leq k(n)$.

\begin{definition} [Domain-constrained Abstraction]
\label{def:abstraction}
Given an MDP $\MdpInitR$ with $\states = D_{v1} \times \dots \times D_{vn}$ and $D_{vi}\subseteq \N$ and cut-off function $k$ with $k(i) \in D_{vi}$ for $i \in [1\twodots n]$.
The abstraction MDP $\shielded{\mdp} = \MdpTupleShielded$ 
has the following components:
\begin{compactitem}
    \item $\states{\tinyshield} =  \shielded{{D_{v1}}} \times \dots \times \shielded{{D_{vn}}}$ with $\shielded{{D_{vi}}} = \{x \mid x \in D_{vi} \text{ and } x \leq k(i)\}$ is the state space,
    \item $\shielded{{s_0}}=(s_0[I \leftarrow k])$ with $I=[1\twodots n]$ is the initial state, 
    
    \item $\shielded{\Act}(s) = \Act(s)$ are   actions available in $s \in \states{\tinyshield}$, and

    \item $\shielded{\pmdp}$ is the probabilistic transition function. 
   For all $I \subseteq [1..n]$ we define the state sets 
   \begin{align*}
   {\states}_{\geq}^I = \{s\in \states \mid&\forall j \in I. ~ s[j] \geq k(j) \land \\
   & \forall j \notin I. ~ s[j] < k(j) \}, \\
   {\states}_{\mathrm{abs}}^I = \{s[I \leftarrow k] \mid &s\in {\states}_{\geq}^I\}.
   \end{align*}
   For all $s \in \states{\tinyshield}$, $a \in \shielded{\Act}(s)$,  $I \subseteq [1..n]$, and $s' \in {\states}_\mathrm{abs}^I$,
   $\shielded{\pmdp}$ is defined by
   \begin{equation}
   \shielded{\pmdp}(s, a)( s') = \smashoperator{\sum_{s'' \in  {\states}_{\geq}^I:
     s''[I \leftarrow k] = s'}} \pmdp(s,a)(s''). \label{eq:prob_abstract}
    \end{equation}

 %
   

\end{compactitem}
\end{definition}

The abstraction lumps together states where values exceed their cut-off points given by the function $k$.
For any index set $I$, ${\states}_{\geq}^I$ contains all states
whose variables indexed in $I$ have values greater than or equal to their cut-off values. The set ${\states}_{abs}^I$ contains all states where the indexed variables have exactly the cut-off value. 
Transitions to states in ${\states}_{abs}^I$ in $\shielded{\mdp}$
are combined transitions to states in ${\states}_{\geq}^I$ in $\mdp$,
therefore we sum their probabilities in \cref{eq:prob_abstract}.
Note, that the empty index set $I$ is a special case, resulting 
in $\shielded{\pmdp}(s,a)(s')  = \pmdp(s,a)(s')$ for all states 
$s$, $s'$  where for all  $v_i \in  s$, $v_j \in s'$ it holds that  $v_i < k(i)$ and $v_j < k(j)$.

\emph{\emph{\textbf{Example.}} 
The values of the variables $n$, $s$, $e$, and $w$ that form the states of $\mdp$ are 
unbounded. To enable synthesis, we lump states in the abstraction $\shielded{\mdp}$ together, where $n$, $s$, $e$, or $w$ exceed specific thresholds.
The domain of $ctr$ is the same for $\mdp$ and the abstraction $\shielded{\mdp}$.
}

%% file: method.tex
In this section, we discuss our adaptive shielding approach for deriving shields that are optimal w.r.t. our knowledge of the environment.

The approach works as follows:
We start from an initial infinite-state MDP $\mdp$ that models the interaction of the environment, the controller, and the shield. From $\mdp$, we derive a finite-state abstraction $\shielded{\mdp}$ and synthesize an initial shield that we use during run-time. The following steps ensure adaptivity at run-time.

    
    \noindent
    {1) \em Continuous updates of $\mdp$:} During run-time, we monitor the environment
    and use the collected observations to update the estimation of the transition function of $\mdp$.
    
    \noindent
    {2) \em Periodical updates of $\shielded{\mdp}$:}
    During run-time, we monitor whether states of $\mdp$ were visited that are abstracted away in the current $\shielded{\mdp}$. 
    Every $t$ time units, we create a new abstraction $\shielded{\mdp}$ from the current 
    model $\mdp$, with additional refinement if necessary.
    From the new  $\shielded{\mdp}$,
        we synthesize a new shield and use it to replace the old one. 

We now detail the individual technical steps to realize our proposed method.

\subsection{Continuous updates of $\mdp$}

As introduced above, we use $\MdpInitR$ to denote the infinite-state MDP underlying the synthesis environment. 
Since the structure of $\mdp$ is known, learning about the environment amounts to estimating $\mdp$'s probabilistic transition function $\pmdp$.

In the following, we follow Chen and Nielsen's formalization of online learning of transition probabilities~\cite{DBLP:conf/icmla/ChenN12}.
For each state $s$ and action $a$, there is a multinomial distribution $\pmdp({\states}^s|s,a,\mathbf{\theta}^{(s,a)})$ parameterized by $\mathbf{\theta}^{(s,a)}$ over the successor states ${\states}^s$ of $s$ that we need to estimate. As is common~\cite{DBLP:conf/icmla/ChenN12,DBLP:conf/cdc/BertuccelliH08}, we express the uncertainty about the transition probability distributions through prior Dirichlet densities $\pmdp(\mathbf{\theta}^{(s,a)})$. We use a Dirichlet distribution for this purpose, since it is the conjugate prior of the multinomial distribution. In general, we use symmetric Dirichlet distributions as prior distributions, as we initially assume all environment changes to be equally likely.

A Dirichlet distribution is parameterized by $(\alpha^{(s,a)}_1, \ldots, \alpha^{(s,a)}_{|{\states}^s|})$, where $\alpha^{(s,a)}_j \in \mathbb{R}^+$. Each $\alpha^{(s,a)}_j$ roughly corresponds to how often a particular successor state is observed after executing action $a$ in state $s$. Given $\pmdp({\states}^s|s,a,\mathbf{\theta}^{(s,a)})$ and a Dirichlet distribution
$\pmdp(\mathbf{\theta}^{(s,a)})$, we can estimate the probability of reaching $s_j$ after executing $a$ in $s$ by $\frac{\alpha^{(s,a)}_j}{\sum_k \alpha^{(s,a)}_k}$.

After observing the successor state $s_j$, we update $\mdp$ by computing the posterior distribution $\pmdp(\mathbf{\theta}^{(s,a)})$ with updated hyperparameters, which is also a Dirichlet distribution. In contrast to Chen and Nielsen~\cite{DBLP:conf/icmla/ChenN12}, we consider a non-stationary environment, that is, the transition probability distributions may change over time. Therefore, we introduce a discounting factor $\lambda< 1$ to discount past observations similarly to Bertuccelli and How~\cite{DBLP:conf/cdc/BertuccelliH08}. We refer to $\lambda$  also as learning rate. Given such a $\lambda$, The update rule of the hyperparameters is defined as follows~\cite{DBLP:conf/cdc/BertuccelliH08}:
\begin{equation*}
    \alpha^{(s,a)}_j = \lambda \alpha^{(s,a)}_j + \delta_{s,a,j},
\end{equation*}
where $\delta_{s,a,j} = 1$ if the observed transition ended in state $s_j$ and $\delta_{s,a,j} = 0$ otherwise. The learning rate $\lambda$ intuitively lets us control how much we value past observations with respect to new observations. Its effect is to keep the estimator's variance non-zero such that new observations have an impact. However, this hinders convergence unless we let $\lambda$ approach $1$ over time~\cite{DBLP:conf/cdc/BertuccelliH08}.

\emph{\emph{\textbf{Example.}}
In our setting, the number of arriving and leaving cars in a single time step is limited, this gives a 
clear structure on the possible successor states. For the sake of simplicity, we assume that the queue sizes 
of each road can change by at most $\pm 1$ and the controller can choose the command $\mathit{NS_c}$ or $\mathit{EW_c}$.
Suppose, the initial state $s = (1,1,2,2,\mathit{NS_c})$ and the shield agrees with the command of the controller
and gives the action $a = \shielded{\mathit{NS}}$.
Let $\lambda$ be $0.9$ and let all prior distributions be initialized with symmetric Dirichlet distributions, where $\alpha^{(s,a)} = (1,\ldots,1)$. There are potentially $162$ successor states for $(s,a)$, thus $\alpha^{(s,a)}$ contains $162$ entries.
After observing the successor state $s' = (0,0,2,2,\mathit{NS_c})$ (the cars in the north-south direction left), we update $\alpha^{(s,a)}$ to $(0.9, \ldots, 0.9, 1.9, 0.9, \ldots, 0.9)$. Each entry is multiplied by $\lambda = 0.9$, except for the entry $\alpha^{(s,a)}_{s'} = 1.9$ corresponding to $s'$, to which we also add $1$. As a result, we estimate the probability of this transition to be $\frac{\alpha^{(s,a)}_{s'}}{\sum_k \alpha^{(s,a)}_k} = \frac{1.9}{161 \cdot 0.9 + 1.9} \approx 0.013$, whereas the probability of each of the other transitions is approximately $0.006$.
}


\subsection{Periodical updates of $\shielded{\mdp}$}

After $t$ time units, we construct a new abstraction $\shielded{\mdp}$ from the current model $\mdp$
using the current cut-off function $k$ via \cref{def:abstraction}. Given the new abstraction $\shielded{\mdp}$, we also compute a new shield and immediately deploy it.

Before creating a new abstraction from the current model $\mdp$, we need to check if 
\emph{abstraction refinement} is needed. Abstraction refinement is needed, if in the last
$t$ time units, states were monitored whose variable values exceeded its  cut-off.
In this case, these states are lumped together in the current abstraction and the cut-off function $k$ needs to be updated.

\begin{definition} [Abstraction Refinement]
Given an MDP $\MdpInitR$ and a cut-off function $k$ defining the maximal value of each variable $v_i \in s$
i.e., $v_1 \leq k(1), \dots , v_n \leq k(n)$.
Given the states ${\states}_t\in \states$ observed in the last $t$ time steps, we formalize abstraction refinement as follows. For all $1 \leq j \leq n$, determine:
\begin{align*}
     {k'}_j &= \max_{s \in {\states}_t}\{ s[j]\}\text{ and} \\
     {k''}_j &= \max({k'}_j, k(j)) \text{ to update $k$ by } \\
    k(j) &\leftarrow {k''}_j.
\end{align*}

\end{definition}

By updating the cut-off function $k$, the state space of the new abstraction $\shielded{\mdp}$ will be refined
based on the last observations of the environment.

\subsection{Detecting Changes}
If the current abstraction is consistent with our current knowledge of the environment dynamics, we may avoid computing a new abstraction and a new shield, thus saving computational resources. \emph{Discrepancy checks} between abstraction $\shielded{\mdp}$ and $\mdp$ can be introduced to avoid unnecessary computations.


In general, we may apply statistical tests for each state-action pair to check if recently observed transitions are consistent with the current abstraction of the environment,
i.e., there are no significant changes of the transition probabilities.
If data is likely to be sparse for most pairs, we propose to apply application-specific checks for changes. 

\emph{\emph{\textbf{Example.}} 
In traffic control, we consider a large state space, thus the data for individual state-action pairs are sparse. Since our goal is to detect changes in the traffic flow, we propose to monitor the average number of cars waiting in each lane at a junction. These numbers correspond to the variables $v_i$ making up the state space. Change detection could be initialized by computing the mean number of cars $\overline{v}_i$ for each $i$ from the first $t$ observations. After that, a standard change detection algorithm, like CUSUM~\cite{CUSUM_original} could be applied.
Whenever a change of $\overline{v}_i$ is detected for some $i$, we can assume that there is a discrepancy between the current abstraction and the current state of the environment. After detecting a change, $\overline{v}_i$  needs to reinitialized for each $i$ using the next $t$ observations. 
}

%% file: experiments.tex
In this section, we evaluate our adaptive shields via simulations with the traffic simulator
``Simulation of Urban MObility'' (\sumo)~\cite{lopez2018microscopic}. 
We use the \sumo C++ API to extract the traffic-light-controlled
intersection’s information and to send orders to change the
traffic light phases.
We implemented the computation of abstraction MDPs in C++. Given abstraction MDPs, we compute shields via value iteration with we the model checker \storm~\cite{hensel2020probabilistic}. 
All simulations were executed on a
desktop computer with a $4$ x $2.70$ GHz Intel Core i$7$-$7500$U CPU, $7.7$ GB of RAM
running Arch Linux.
The simulation results are presented to show the effectiveness of
our approach. The source code and several \textbf{videos of shielding in action} can be found on \url{https://adaptiveshielding.xyz}.

\textbf{Scenario 1: Changing traffic density on a single crossing.}
In the first experiment, we use the crossing illustrated in Fig.~\ref{fig:states_M} for simulations.
The intersection is composed of four roads, and every road has two lanes:
the right lane allows right turns,
and the left lane allows the left-turn and through vehicles. During the simulations, we measure the waiting time of cars on the road.

The simulations of the experiment last for $6000$ time steps (1 time step corresponds to 1 sec).
At the beginning of a simulation, the average vehicle arrival rate is uniformly distributed over all lanes with
an arrival period of $1.5$ cars per second and captures the normal traffic density at this crossing.
After $500$ steps, the traffic density on the east-to-west
road increases to $65\%$ of all the arriving cars (e.g., due to an accident or a construction site).

\emph{The controller.}
In each simulation, the controller uses a fixed stage 
duration of $20$ seconds for both 
directions.

\emph{The shield.} We compute adaptive shields for this junction.
Its MDP representation is illustrated on the right in Fig.~\ref{fig:states_M}.
Each state of the model $\mdp$ is a 5-tuple $(n,s,e,w,ctr)$ composed of integers.
The variables $n$, $s$, $e$, and $w$ represent 
the number of waiting cars in each direction and $ctr\in\{NS_C,EW_C\}$ is the issued command of the controller.
We use the knowledge about the initial traffic density and the signal's stage duration of the controller for an initial estimate of the transition probabilities of $\mdp$.
To construct the initial abstraction $\shielded{\mdp}$, we use the cut-off function $k(i)=3$ for $i\in[1\twodots4]$.
As performance measure we charge an abstract state $s=(v_1,v_2,v_3,v_4,ctr)$ with a cost of
$c_1 = |\max(v_1,v_2)-\max(v_3,v_4)|$, thus the shield aims to balance the
number of waiting cars per direction. 
Interfering, i.e., altering the controller action, incurs a constant cost value $c_2$ to the shield.
The factor $\gamma=0.5$ is chosen to weight the costs $c_1$ and $c_2$. 

\emph{Adaptive updates.}
During run-time, we continuously monitor the traffic density and 
update the probability distributions of $\mdp$ with a learning rate $\lambda$. Additionally, we update the cut-off function $k$ if we observe states, whose values exceed their respective cut-off values. Every $t=500$ time steps, after an initial setup time of $1000$ steps, we compute a new abstraction $\shielded{\mdp}$ and a new shield based on $\shielded{\mdp}$, which we immediately deploy.


\emph{Results.}
Throughout all our experiments, we use a unified measure of performance: the total waiting time of the cars in the city accumulated over the 
last 100 time steps.
Our assumption is that minimizing this measure is the main objective of the designer of the traffic
light system for the city.
We show the effect of the shield-interference cost w.r.t. the concrete performance.
We experimented with shields with
a learning rate of $\lambda=0.3$ and
different fixed interference costs $c_2 \in \{5,7,9\}$.
Note that the same effect could be achieved by varying $\gamma$.

\hspace{5mm}
\begin{figure}
    \centering
    \includegraphics{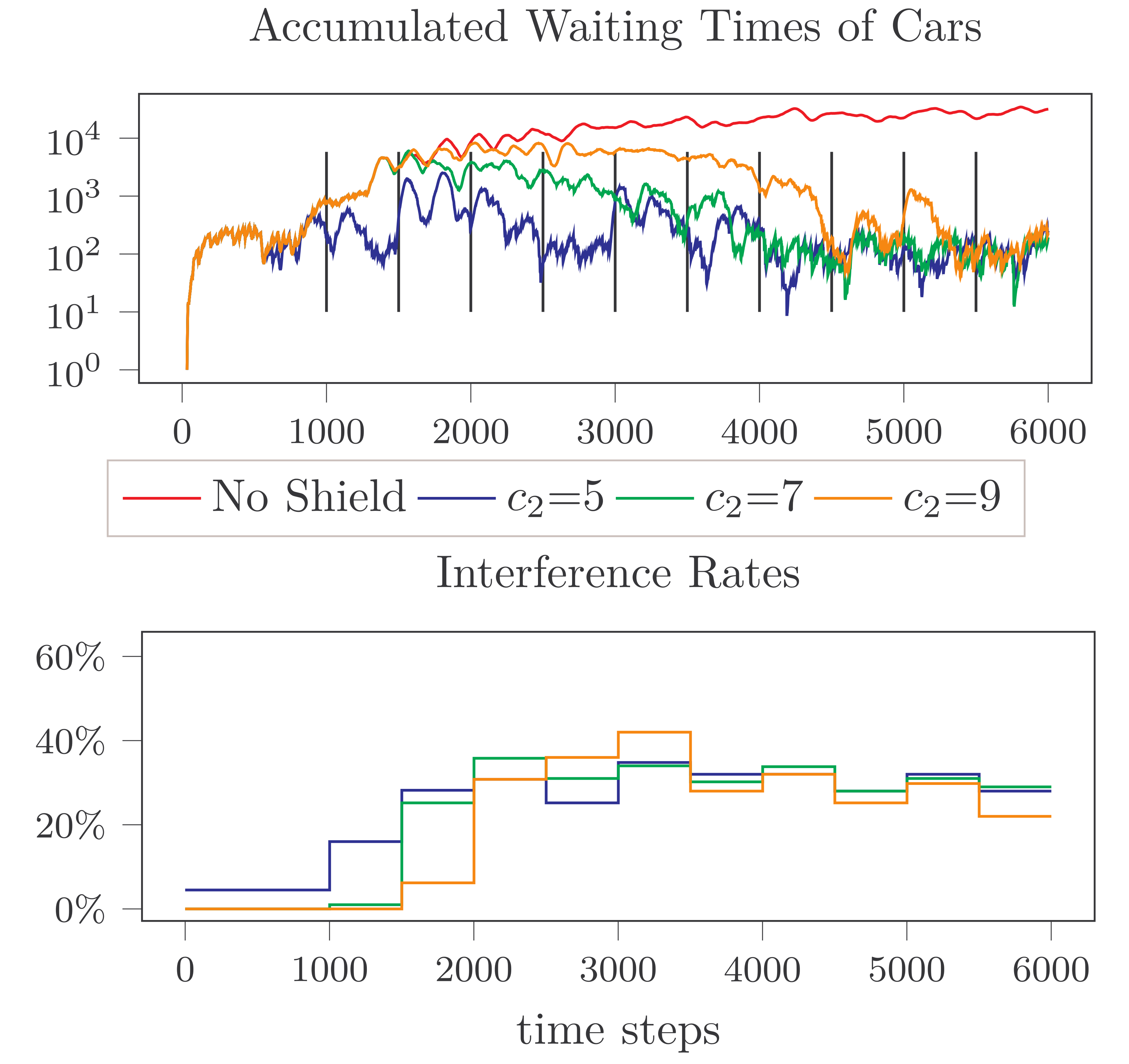}
    \caption{Results for Scenario 1: Shielding a single intersection.}
    \label{fig:baseExperimentPlot}
\end{figure}

\begin{table}[t]
\centering
\caption{Updates of the cut-off function $k$ for different shields.}
\label{tab:baseExperimentTable}
\begin{tabular}{|c|c|c|c|}
\hline 
\multicolumn{1}{|c|}{\textbf{\begin{tabular}[c]{@{}c@{}}time-\\ step\end{tabular}}}  &
\multicolumn{1}{|c|}{\textbf{\begin{tabular}[c]{@{}c@{}}$k$ for shield\\ with \textbf{$c_2=5$}\end{tabular}}}   & 
\multicolumn{1}{|c|}{\textbf{\begin{tabular}[c]{@{}c@{}}$k$ for shield\\ with \textbf{$c_2=7$}\end{tabular}}}   & 
\multicolumn{1}{|c|}{\textbf{\begin{tabular}[c]{@{}c@{}}$k$ for shield\\ with \textbf{$c_2=9$}\end{tabular}}}   
 \\\hline  
0       & (3,3,3,3) & (3,3,3,3) & (3,3,3,3)   \\\hline 
1000    & (4,4,4,4) & (4,4,4,4) & (4,4,4,4)   \\\hline 
1500    & (5,4,5,4) & (5,4,5,4) & (5,4,5,4)   \\\hline 
2000    & (5,4,6,4) & (5,4,6,4) & (5,4,6,4)   \\\hline 
2500    & (5,5,7,4) & (6,5,7,4) & (5,5,7,4)   \\\hline 
3000    & (6,6,8,4) & (7,6,8,4) & (6,6,8,4)   \\\hline 
3500    & (7,6,9,4) & (8,7,9,4) & (7,7,9,4)   \\\hline
4000    & (8,6,10,4)& (9,7,10,4)& (8,8,10,4)  \\\hline
4500    & (8,7,11,4)& (9,7,11,4)& (9,9,11,4)  \\\hline
5000    & (8,7,11,4)& (9,7,11,4)& (10,9,12,4) \\\hline
5500    & (8,7,11,4)& (9,7,11,4)& (10,9,12,4) \\\hline
6000    & (8,7,11,4)& (9,7,11,4)& (10,9,13,4) \\\hline
6500    & (8,7,11,4)& (9,7,11,4)& (10,9,13,4) \\\hline
7000    & (8,7,11,4)& (9,7,11,4)& (10,9,14,4) \\\hline
7500    & (8,7,11,4)& (9,7,11,4)& (10,9,14,4) \\\hline
\end{tabular}
\end{table}

Fig.~$2$ shows the effect of the shields on the performance as well as the according ratios of times where the shields intervene. 
Table~\ref{tab:baseExperimentTable} summarizes the updates of the cut-off functions of the individual shields.
In Fig.~\ref{fig:baseExperimentPlot} the red line shows the performance of the unshielded controller, which is good until step 500 when the traffic density shifts to one direction. From this time step on the controller has to face unprecedented situations and therefore drops in performance. The remaining lines show the effect of using shields with $c_2\in\{5,7,9\}$.   
We observe that all shields are able to reduce the total waiting time of the cars
by orders of magnitude. For larger values of $c_2$, interference
is too costly in the beginning. Several updates of the cut-off function are needed until the abstraction captures the number of waiting cars with sufficient accuracy. With increasing accuracy the shield starts to intervene and the waiting times improve. 
After several refinement steps, all shields achieve the same level of 
performance with identical interference rates.

In this setting, we used a learning rate of $\lambda=0.3$. We conducted the same experiment with different values of $\lambda$ and noticed only minor
differences in the accumulated waiting times. For larger $\lambda$, the accumulated waiting time dropped a bit sooner.
However, we noticed that adjusting the costs for interference
has more impact than changing the learning rate. 
More experiments in different settings are necessary to give a general statement.

\begin{figure}[t]
\center
\includegraphics[width=7.2cm]{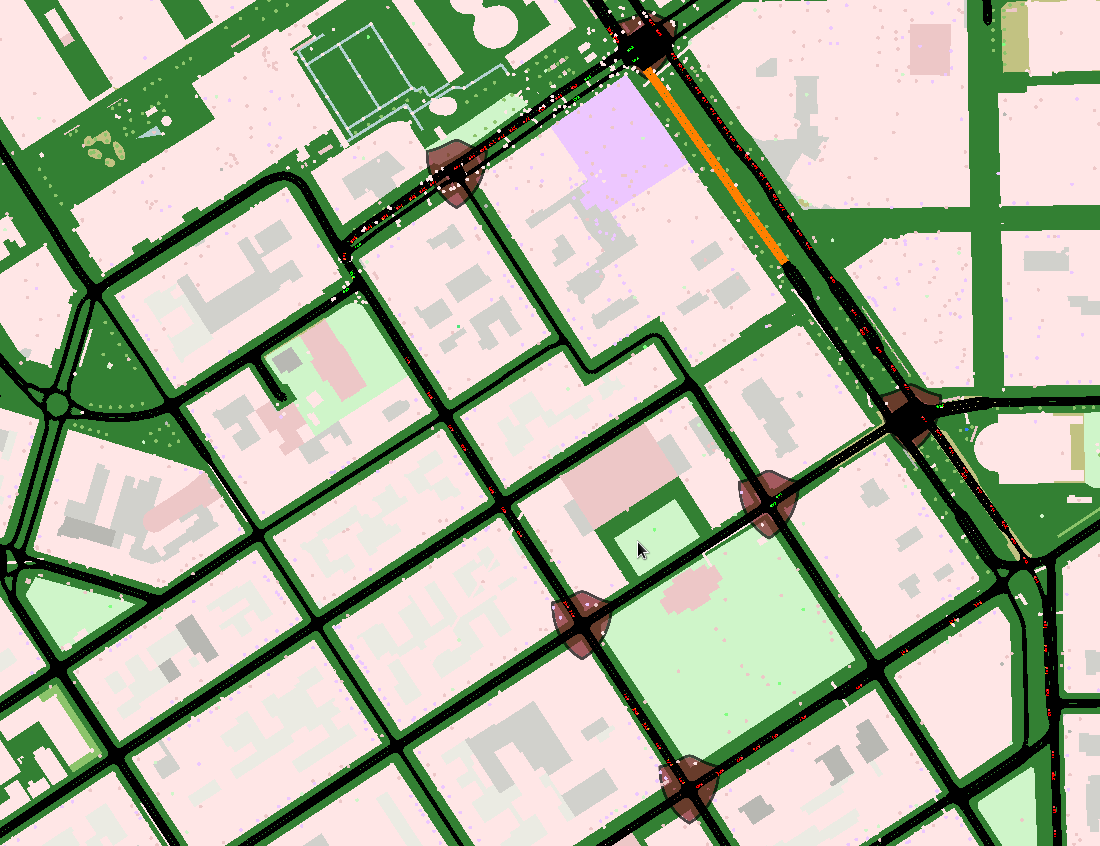}
\label{fig:helsinkiWithShields}
\caption{
 \sumo representation of a selected area of the Helsinki traffic network showing $6$ shielded junctions (shaded red).
}
\end{figure}

\begin{figure}[t]
    \centering
    \includegraphics{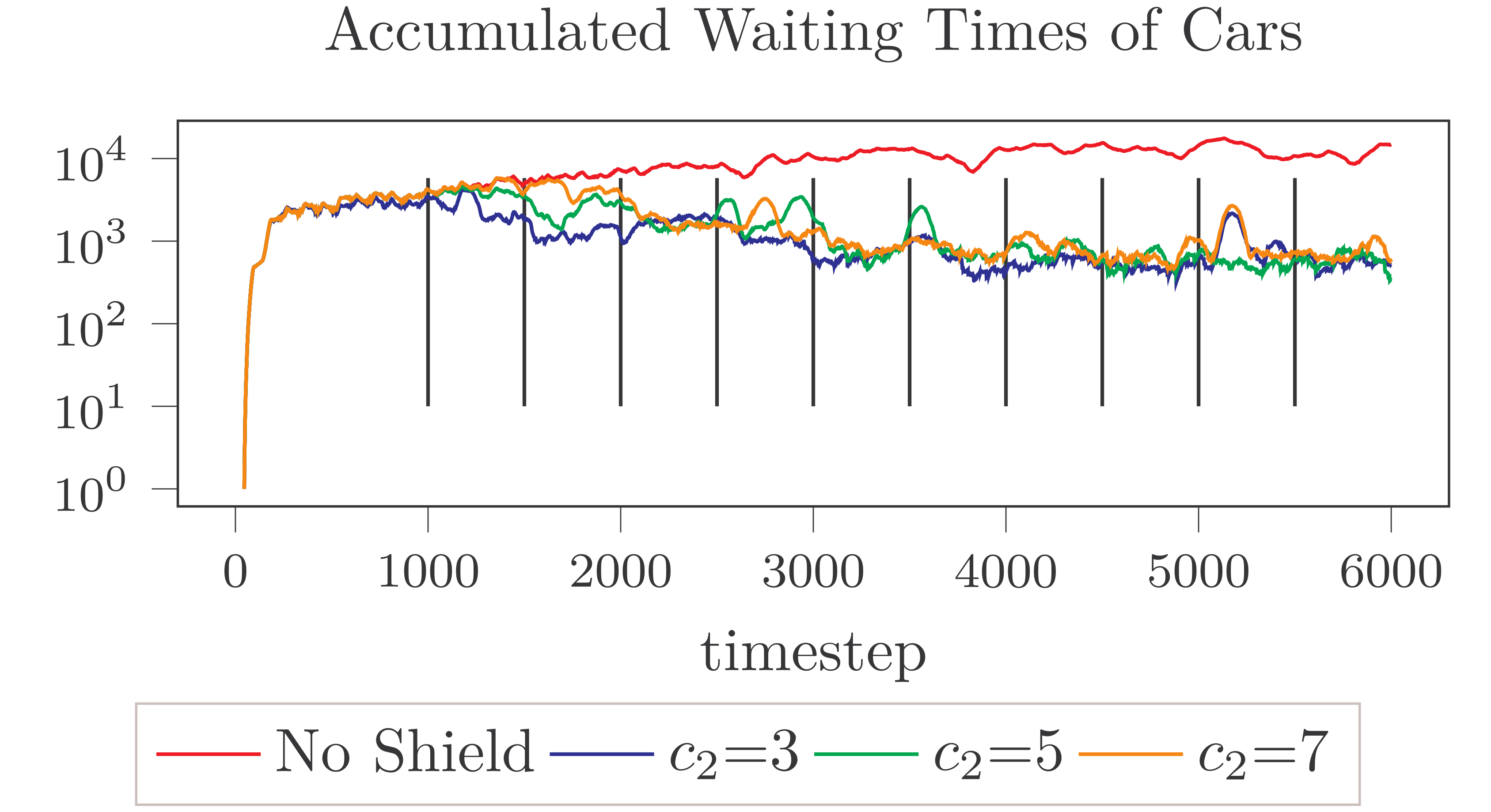}
    \caption{Results for Scenario 2: Changing traffic in Helsinki.}
    \label{fig:baseHelsinkiExperimentPlot}
\end{figure}
\textbf{Scenario 2: Changing traffic density on a road network.}
We tested our adaptive shields by using a part of the urban traffic network of Helsinki and shielded 6 selected intersections as shown in Fig.~$3$. 
The main traffic flow across the city is on the highway. We use a traffic
light controller that is optimized for that scenario.
In the simulation, we block a section of the highway in the north-south direction
from time step 1000 to 6000.
This causes the cars to take a detour on the smaller roads before 
getting back on the highway.

\emph{The shields.} 
Due to the detour of the cars that leave the highway, the traffic density on
two junctions on the highway and four junctions on side roads changes drastically.
We place a local shield at each of this six crossings. 
Note that the fact that we synthesize local shields makes our approach highly scalable
and we could easily place a shield at any junction in the city.
We compute for each crossing a model based on the usual traffic density
and the available controller actions and synthesize initial shields.
The cost functions and parameters are the same as in Scenario 1. 

\emph{Results.} 
Fig.~\ref{fig:baseHelsinkiExperimentPlot} shows the effect of the shields on the accumulated waiting time of all cars
in the city. The results of Scenario 2 substantiate the previous results.
The controller without a shield (red line) performed well initially 
until the traffic situation changed and traffic started to congest due to
non-adaptive traffic control. The shields adapted to the new traffic density quickly
and reduced the waiting times by orders of magnitude. 

\textbf{Scenario 3: Prioritizing Public Transport}
This scenario demonstrates that shields can be used to add functionality to
an existing controller. 
Especially when designing learned controllers, shields can be seen as a tool for designers to optimize a secondary objective and to keep the reward structure of the learning agent simple.

\emph{The shield.}
The model $\mdp$ over which the shield is constructed slightly differs from the one
used in the other experiments. 
The state space is likewise composed of 5-tuples  $s=(n,s,e,w,ctr)$, but this time
the variables $n$, $s$, $e$, and $w$ represent the total waiting time of
all buses on the corresponding roads. For example, the state $(0, 21, 0, 0, EW_C)$ represents that buses are waiting in the South queue for a total time of $21$ seconds.
Outgoing edges take the frequencies of arriving buses into account. 
As a performance measure, we assign a cost to each abstract state, which is equal to the difference in the bus waiting times per direction. All other parameters a kept the same. 

\begin{figure}[!t]
    \centering
    \includegraphics{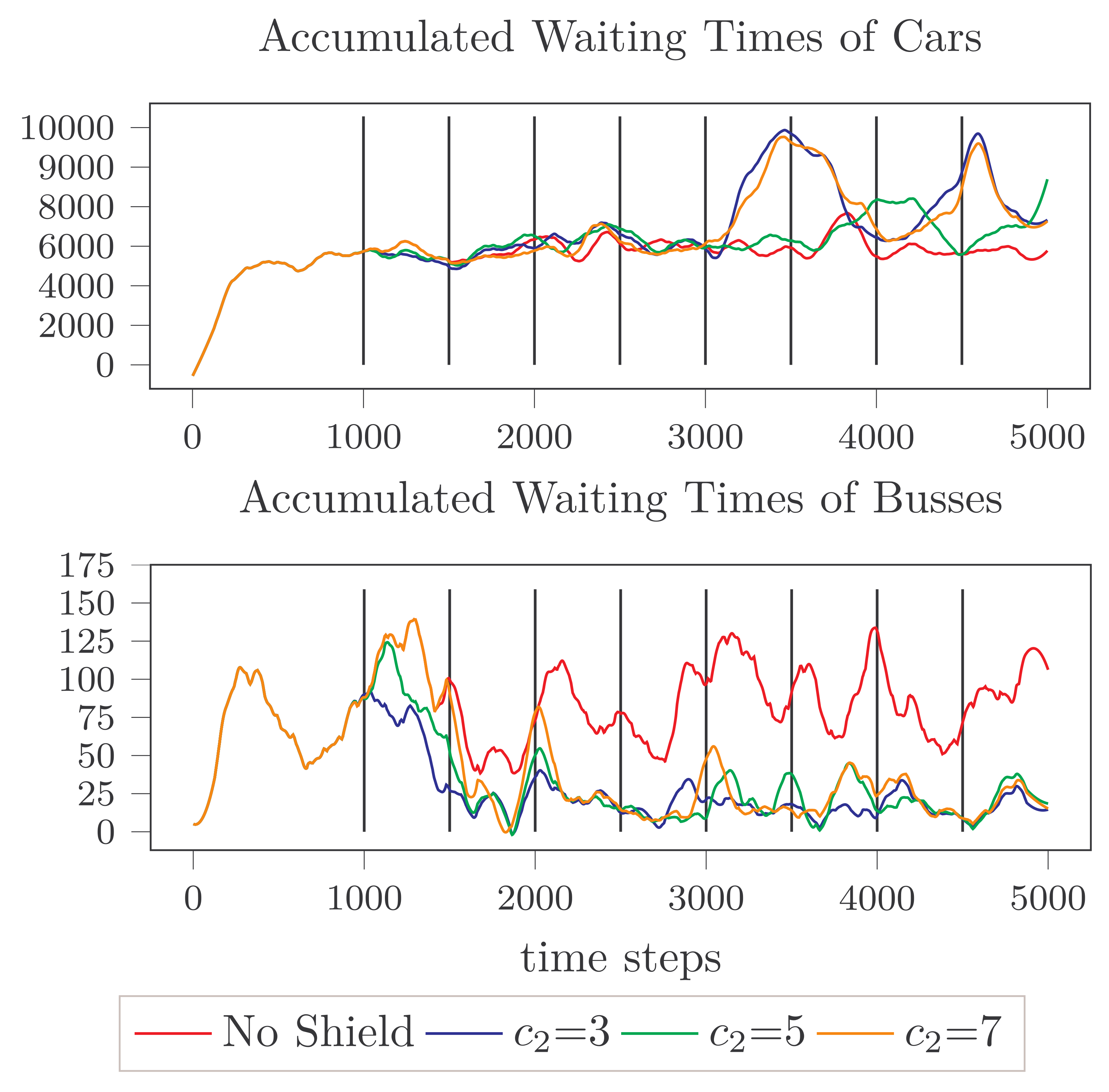}
    \caption{Results for Scenario 3: Influence on normal traffic.Results for Scenario 3: Prioritizing public transport over normal traffic.}
    \label{fig:busExperimentCarsPlot}
\end{figure}

\emph{Results.}
In Fig.~\ref{fig:busExperimentCarsPlot} we depict the total waiting time of all vehicles and only busses as a function of the interference cost $c_2$. 
We observe the predicted behavior. The interference of the shield
improves the waiting time of the buses at the expense of the general waiting time.

\input{cavComparison}

\textbf{Synthesis times.}
Table~\ref{tab:cavComparison} presents the shield-synthesis times of 
our implementation in comparison to the time required by Avni et al.'s implementation~\cite{AvniBCHKP19}.

We compute the shields for an intersection of four roads, with a uniform distribution of arriving cars, and a uniform cut-off value $k$ for all lanes. The first column of Table~\ref{tab:cavComparison} lists specific cut-off values. The second column lists the synthesis times of Avni et al.'s implementation 
and the last column lists the synthesis times of our implementation.
The timeout is set to $900$ seconds. 
Avni et al. treated the controller adversarially, thus computing a shield 
requires solving a stochastic mean-payoff game with 2 players (2.5-player game).
The authors implemented a strategy iteration
algorithm~\cite{DBLP:books/daglib/0007403} which starts with an initial policy (strategy) that is iteratively improved. 
In our framework, we model the controller probabilistically, thus computing a shield
requires solving an MDP with mean-payoff objective. In other words, we need to compute one strategy only once, whereas 2.5-player games require the computation of two strategies, one for the shield and one for the adversarial controller, in an alternating manner. While solving 2.5-player games is unlikely to be NP-hard~\cite{AvniBCHKP19}, Table~\ref{tab:cavComparison} shows that it requires substantially more time than computing a strategy for an MDP. We also compared the shielding performance 
and achieved the same level of performance with our modeling approach.

%% file: cavComparison.tex
\begin{table}[!ht]
\centering
\caption{Comparison of Synthesis Times}
\label{tab:cavComparison}
\begin{tabular}{|c|c|c|}
\hline 

\multicolumn{1}{|c|}{\begin{tabular}[c]{@{}c@{}}$k$\\~\end{tabular}}  &
\multicolumn{1}{|c|}{\begin{tabular}[c]{@{}c@{}}synthesis time [sec]\\ shields from $2.5$-player games~\cite{AvniBCHKP19}\end{tabular}}   & 
\multicolumn{1}{|c|}{\begin{tabular}[c]{@{}c@{}}synthesis time [sec]\\ shields from MDPs \end{tabular}} \\ [0.5ex] 
\hline  

4 & 1.5 & 0.063   \\ \hline
5 & 4.7 & 0.064   \\ \hline
6 & 14.3& 0.079   \\ \hline
7 & 85  & 0.099   \\ \hline
8 & 377 & 0.124   \\ \hline
9 & 632 & 0.158   \\ \hline
10&  -  & 0.224   \\ \hline
15&  -  & 0.993   \\ \hline
20&  -  & 2.997   \\ \hline
25&  -  & 7.380    \\ \hline
30&  -  & 15.481 \\ \hline

\end{tabular}
\end{table}

%% file: conclusion.tex
We presented adaptive shielding, an approach for adaptive control through local changes of the global control scheme at run-time. 
Shields can be used to ensure that safety and performance requirements are met, or to add functionality, while being minimally interfering with the global controller. In this work we focus on shields in a changing and uncertain environment.
As underlying model for the shield computation we use MDPs with costs. Adaptivitiy is
ensured by regular updates and refinement steps of this model based on observations of the environment during run-time.
We illustrate our approach using experiments on urban traffic control with changing traffic patterns. Our experiments show that shielding enables adaptation to changing traffic flow, which, e.g., may occur due to accidents or traffic jams. 

For future work, we will examine different ways to learn from observations of the environment. For instance, instead of estimating multinomial distributions for the transition probabilities, we may consider estimating binomial distributions for the probability of individual transitions. Zhao et al.~\cite{interval_MDPs_ASE2020} presented such an approach for interval Markov chains. The authors also presented change-point detection for interval Markov detection. 
We intent to explore different methods for change-point detection to detect changes in the traffic flow. Among others, we want to explore an adaption of the technique of Zhao et al..
Finally, we will apply our adaptive shields on controllers trained via state-of-the-art 
deep reinforcement learning and investigate the performance improvements.